# Tethered Balloon Technology for Green Communication in Smart Cities and Healthy Environment

1st S. H. Alsamhi
Dept. of Electronics Engineering
IBB University, Yemen and Aligarh Muslim University, India
Ibb, Yemen
s.alsamhi.rs.ece@iitbhu.ac.in

2nd Mohd Samar Ansari, IEEE Senior Member
Software Research Institute Athlone
Institute of Technology Athlone
Ireland
mdsamar@gmail.com

3th Liang Zhao
School of Computer Science
Shenyang Aerospace university
Shenyang, China
lzhao@sau.edu.cn

4th Sau Nguyen Van
Biomedical of Information Technology
Shenzhen Insitutes of Advanced Technology, UCAS
Shenzhen, China
saunv.edu@gmail.com

5th S. K. Gupta
School of Electronics and Communication Engineering Shri Mata Vaishno Devi University
Katra, India
sachin.rs.eee@iitbhu.ac.in

6th Amr A. Alammari, IEEE Member
Department of Electronics Engineering
Aligarh Muslim University
Aligarh, India
amr.rs@amu.ac.in

7th Akram Hatem Saber
Department of Electronics Engineering
Aligarh Muslim University
Aligarh, India
alasmr.2a@gmail.com

8th Mohammed Y.A.M.Hebah
Department of Electronics Engineering Aligarh Muslim University
Aligarh, India
moha.hebah@gmail.com

9th Marwan Ahmed Abdullah Alasali
Department of Electrical Engineering
Aligarh Muslim University
Aligarh, India
ma734608@gmail.com

10th Hasan Mohsen Aljabali
Department of Electronics Engineering
Aligarh Muslim University
Aligarh, India
hasanaljabali45@gmail.com

11th Mohd Najim
Department of Electrical and Electronics Engineering
University of Jaddah
Jaddah, Saudi Arabia
mohdnajim.iitr@gmail.com

12th Ashutosh Srivastava
Department of Electrical Engineering
IIT (BHU)
Varanasi, India
ashutosh.rs.eee@iitbhu.ac.in

*Abstract*—The development and adopting of advanced communication technologies provide mobile users more convenience to connect any wireless network anytime and anywhere. Therefore, a large number of base stations (BS) are demanded keeping users connectivity, enhancing network capacity, and guarantee a sustained users Quality of Experiences (QoS). However, increasing the number of BS leads to an increase in the ecological ad radiation hazards. In order to green communication, many factors should be taken into consideration, i.e., saving energy, guarantee QoS, and reducing pollution hazards. Therefore, we propose tethered balloon technology that can replace a large number of BS and reduce ecological and radiation hazards due to its high altitude and feasible green and healthy broadband communication. The main contribution of this paper is to deploy tethered balloon technology at different altitude and measure the power density. Furthermore, we evaluate the measurement of power density from different height of tethered balloon comparison with traditional wireless communication technologies. The simulation results showed that tethered balloon technology can deliver green communication effectively and efficiently without any hazardous impacts.

## I. INTRODUCTION

Tremendous growing in wireless communication technologies provide mobile users freedom to inter operate with any existing communication technologies. However, the development of such technologies make people demand for better and high QoS and low cost. To fulfill the mobile users desires, many BS are required to be deployed for ensuring sustainable user Quality of Experience (QoE). Increasing number of BS leads to increase several impacts such as radiation hazards due to the excessive electromagnetic radiation [1], increase the running cost in rural area, and increase the carbon footprint in atmosphere due to the excessive burning of diesel in BS.

Electromagnetic (EM) spectrum has consistently shown the health hazard from the radio frequency range (RFR), which leads to concern [2], DNA damage and so on [3]. Therefore, people who are living more closer to towers are overexposed to EM radiation [3], and females having more complaints than males [2]. So, the location of towers must be far away from human habitats to minimize the effects of EM and RFR. Furthermore, fuel-burning affects the human living environment. In many countries, the cell towers run on diesel, which leads to an increase in $CO_2$ emission [2]. $CO_2$ the emission has estimated around 10000000 tons every year [4]. For more evident, CO2 emission has reached $1.7 * 10^9$ tone only from Information and communications technologies (ICT) which contributes between 2%-10% of global $CO_2$ [4]. Reducing carbon by using solar energy is not a solution for the green environment because the effects of radiation still harm human beings, the environment, and society. Therefore, space technologies play a crucial role in protecting and enhancing our society from wireless communication effects and hazardous. Furthermore, space technology can deliver broadband services to a large coverage area with healthy and green communication. However, delivering communication services is very costly.

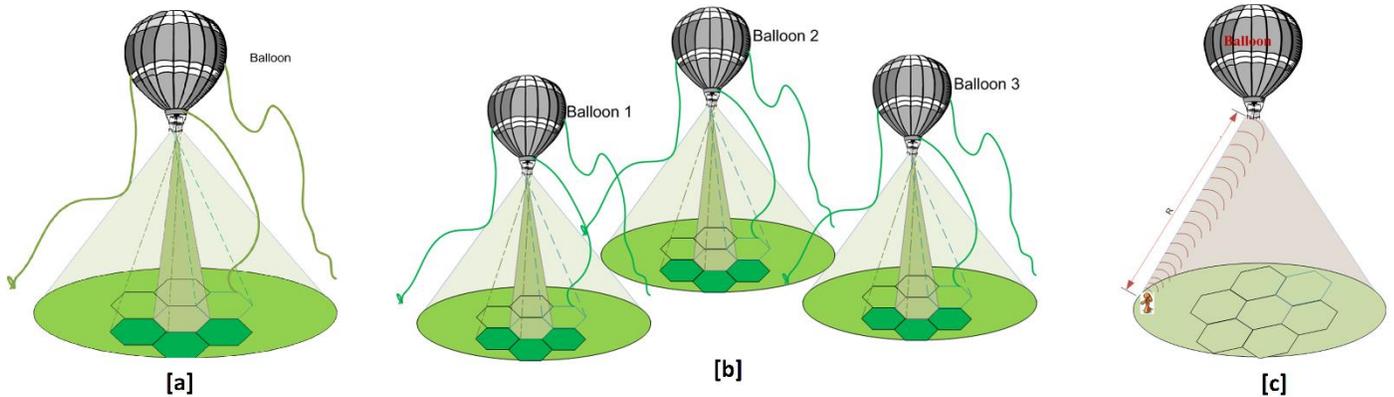

Fig. 1: (a) Single tethered balloon; (b) Multiple tethered balloon; (c) Power density from tethered balloon.

Through the appropriated understanding of the health hazards of radiation, carbon emission, and power consumption demanded, we found tethered balloon technology will represent the critical solution for enhancing the healthy environment and green communication effectively and efficiently. Many existing kinds of research have emerged on tethered balloon technology network architecture, coverage area, and its application for disaster mitigation and recovery [1], [5]–[13], but none of these studies focused on the application of tethered balloon for green communication in smart cities for improving the health quality. These researches are exciting, but they have applied for limited applications. First of all, the authors [8] summarized the enabling technologies for greening Internet of Things (IoT). The authors of [1], [6] discussed tethered balloon network architecture and the coverage for delivering broadband communication services. However, the authors of [7], [9], [13] explored that tethered balloon represented the efficient solution for disaster prevention, mitigation, management and recovery. However, the studies [10], [11] focus on the classification of environments and health hazards while the effects of exposure to EMFs on the body and cells discussed in [11]. Therefore, we propose a tethered balloon as a robust technology that can improve human life quality and make it more comfortable and healthy. The rest of the paper is organized as follows. Section II shows the advantages and applications of tethered balloon technology. The proposed network architecture is presented in Section III. In Section IV, network design is proposed, and performance analysis is discussed in Section V. Finally, results and conclusions are drawn in Sections VI and VII, respectively.

## II. TETHERED BALLOON TECHNOLOGY

Tethered balloon is one of the low altitude platform family (LAP), which operates in troposphere 200-440m [6], [10]. Zeppelin launched a craft in 2000, followed by many projects which focused on the design of tethered balloon for various applications from different altitudes [14]. The authors of [15] discussed machine learning and the optimal propagation models from different altitude. Then, in 2003, CAPANINA project focused on providing broadband services [16]. Internet connectivity via the balloon network was proposed by the Loon project. In 2001, Gawale et al. [17] aimed to develop an airship for various applications such as surveillance, disaster management, and advertisement. Tethered balloon established a long-distance wireless communication link to provide Internet connectivity in rural areas [16]. Furthermore, Chopra et al. [4] attempted to provide telecommunication and broadband services in remote and hilly areas.

Hence, Tethered balloon technology has the capability to overcome the challenges which appear in traditional cellular systems such as high-power consumption, limited coverage area, and lack of line of sight (LoS). Furthermore, it carries the the potential of mitigating the health risk due to radiation exposure and environmental effects that are linked with terrestrial mobile BS. Therefore, tethered balloon technology has several advantages over terrestrial communication such as low cost of launch and operation, rapid deployment, large coverage area, low propagation delay, easily reconfigurable, and LoS [10], [18].

Tethered balloon can be used for different applications such as disaster recovery, observation and meteorological usage for being able to gather data [19], [20], delivering power [21], monitoring civilian activities and delivering wireless telecommunication services efficiently and enables large coverage area as well as support wired and wireless communications in the disaster area and healthy recovery [22].

## III. TETHERED BALLOON FOR BROADBAND COMMUNICATION

Tethered balloon can cover a large area and replace a massive number of terrestrial towers in urban areas. The coverage area can be extended via multi-balloons depends on many parameters such as radiation power, the height of the antenna, environment, and propagation path. The value of

allowable LoS is also substituted in Hata model [6], [7] with 'correction factor' of a small city taken as:

$$\text{Correct Factor} = (1.1 \log_{10}(f_c) - 0.7)h_{re}$$
$$-(1.56 \log_{10}(f_c) - 0.8) \quad (1)$$

$$P_l = 69.55 + 26.16 \log_{10}(f_c) - 13.82 \log_{10}(h_{te})$$
$$-a(h_{re}) + (44.9 - 6.55 \log_{10}(h_{te})) \log_{10} D \quad (2)$$

The variable $D$ in Hata model equation determines the cell radius of the coverage area (km), and $f_c$ is frequency, $h_{re}$ represents height of the mobile antenna, $h_{te}$ is BS antenna height, $a(h_{re})$ represent the correction factor, $P_l$ is the path loss. A single tethered balloon is envisaged for providing communication services for special events such as disaster recovery, rapid broadband employment over highly populated cited as shown in Fig. 1(a).

*A. Coverage Extension Using Multi-Balloon*

Multi-balloons constellations can increase the capacity by exploiting the directionality of the fixed user antenna. Configuration of tethered balloon leads to provide resilience and unique diversity to end users receiving antennas for improved service availability. In the case of multi-balloon constellations, they can be interconnected via ground stations or via inner balloon link (IBL), which represents an additional network interface between multi-balloons. IBLs can be radio or optical, thus arbitrarily extending the system coverage. In fact, the coverage area is divided into multiple cells for increasing the capacity. Tethered balloon constellations architecture is depicted in Fig. 1(b). Worldwide interoperability for Microwave Access (WiMAX) BS should be taken into the payload of the balloon to deliver broadband services [23], [24].

## IV. HAZARDS

Our society and environment should be given priority during the design of wireless technologies equipment with considering their long-term adverse effects. Threats posed by the introduction of such devices can be classified as environmental hazard and health hazards [10]. Recently, the EM radiations are emitted from phone towers estimated beyond safety levels. Radiation is a form of energy on the move and consists of electric waves and magnetic waves which moving together through space at the speed of light. Various neuro-behavioral changes were observed in the inhabitants living near the towers. Also, diseases had been reported among individuals such as headaches, dizziness, memory changes, sleep disorder.

*A. Electromagnetic Radiation Hazards from BS*

The electrical and magnetic equipment produce invisible electric and magnetic fields (EMFs) and electromagnetic radiation (EMR) in which continuously attacks the human body affecting its bio-field. EMR is a form of energy emitted which attacks human biology depending on the radiated power density as well as the distance from the transmitter [25]. It is measured by the specific absorption rate (SAR in W/kg) and classified into two forms: non-ionizing radiation and ionizing radiation [26]. Non-ionizing radiations do not have enough energy to break apart atoms and molecules and turn them into ions. However, electromagnetic radiation could be considered as ionizing, if the energy associated with a photon of radiation has larger energy than ionizing energies of atoms. For example, X-rays and gamma rays are taken to be ionizing. People who are living within 50 to 300 m radius ius are in the highest radiation zone (dark/blue) and are more prone to ill-effects of electromagnetic radiation, as shown in Fig. 2.

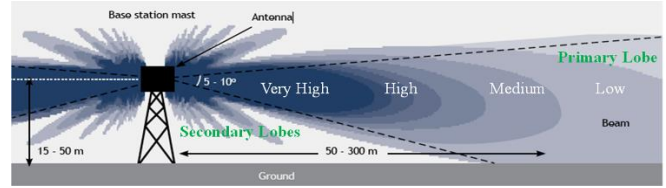

Fig. 2: EMR levels from the BS

*B. Biological Effects*

Currently, active radiation can induce capable of delivering an electric shock to persons or animals. Non-ionizing radiation is associated with two significant potential hazards: electrical and biological. It does not damage the genetic material (DNA) in molecules, directly. Therefore, it cannot cause cancer or any other illness for humanity. X-ray and gamma rays are forms of ionizing radiation. Mainly, these can increase one's risk of cancer, congenital disabilities, and genetic defects through DNA mutations resulting from the atom and molecule ionization. EMR and EMF create an artificial stress situation in the bio-system, which, in turn, affects the metabolism as well as hormone production.

The biological effects of EMR, severe consequences on the health such as DNA damage, increased risk of cancer, breaching of Blood-Brain Barrier, irreversible infertility, effects on the eye, ear, skin, sleeping disorders [27]. Power radiation is defined as the amount of power per unit area in the radiated electromagnetic field at any distance from the antenna. It has uniform power density in a specific direction. Near-field is a region of the electromagnetic field close to the antenna, which lies between the reactive field and far-field [28].

$$n_f = 2L^2/\lambda \quad (3)$$

where $n_f$ is near field, $L$ is the dimension of antenna and $\lambda$ is wavelength. The power density value is calculated within the range of near field, substituting the distance value between two antennas (transmit and receive antenna) in the power density equation which given in [28].

$$P_d = \frac{P_t G_t}{4\pi R^2} \quad (4)$$

where $P_d$ is power density, $P_t$ is transmitted power, and $G_t$ is transmitted antenna gain. In the case of terrestrial wireless communication, for $P_t = 20W$; $G_t = 17dB$; $P_d$ is given for various distance $R$ as Table. I.

TABLE I: Variation of Power Density with Distance

| Distance $R$ (m) | Power density $P_d$ $\left(\frac{w}{m^2}\right)$ |
|---|---|
| **10** | 0.796 |
| **100** | 0.008 |
| **500** | 0.000318 |

On the other hand, $P_d$ from tethered balloon is calculated by using (4), and $R$ is considered as the square root of tethered balloon altitude plus the distance in the near field as shown below.

$$R = \sqrt{a_t^2 + d_{nf}^2} \quad (5)$$

where $a_t$ is the altitude of tethered balloon and $d_{nf}$ is the near field distance. Increasing the distance between the radio station, and the receiver leads to decrease power density. Therefore, radiation power from tethered balloon will not harm or cause hazard to our environment and society in the particular coverage area, as shown in Fig. 1(c). On the other hand, decreasing the distance between the tethered balloon and user on the ground will lead to harm but it does not affect that much as in the traditional wireless communication. Therefore, power density is calculated by using (6), where $R$ represents the distance between tethered balloon and users on the ground.

$$P_d = \frac{P_t G_t}{4\pi R^2} \quad (6)$$

Also, the rms value of the E-field can be calculated as:

$$E = \frac{\sqrt{30 P_t G_t}}{R} \quad (7)$$

The received power at the particular point is:

$$P_r = \frac{P_t G_t G_r \lambda^2}{(4\pi R)^2} \quad (8)$$

### C. Thermal Effects of EMF

The radiation effects are divided into thermal and nonthermal effects. Many studies have been reported that nonthermal effects are more harmful than thermal effects by 3 to 4 times [29]. The adverse health effects are associated with RF energy exposure in the frequency range from 100 kHz to 300 GHz, which is related to the thermal effects.

The harmful effects of radio BS become evident. EMF above 100 KHz can lead to significant absorption of energy and increasing the temperature [5] and it is called thermal effects of EMF. Hence, the effects of exposure to EMFs on the body and cells depend on the EMF frequency and strength [11]. Subsequently, organizations such as WHO, IEEE, and ICNIRP, has been taken into account the risk of thermal effects of high energy radiation, determined a threshold exposure level basis of a health risk assessment of the scientific data below health hazards have been found.

### D. Ecological Hazards

Huawei has reported that carbon emissions of telecommunication equipment is mainly from BSs [30]. According to the international energy agency (IEA), carbon emissions have increased by about 80%. The solution is to have more numbers of cell towers with lesser transmitted power. Lesser transmitted power leads to reduce the heating effects. However, the reduction in the transmitted power for the above solutions will increase the installation and maintenance cost. The operators all over the world are claiming that there are no radiation health hazards. It results in complaining the subscribers of network problems. To avoid these problems solar panels can be used. So, high power diesel generators will not be required. It will reduce carbon emission. Solar systems have a tremendous contribution to reducing $CO_2$ emission as it does not consume fuel. Furthermore, another advantage is a lower cost of operation and maintenance when compared with a diesel system, for the same power delivery.

### V. PROPOSED TECHNOLOGY

Traditional wireless communication technologies have several impacts on our life style and environments. Environment impacts are including increasing the $CO_2$ emission of atmosphere, generators pollution and water resource pollution the result from powering the radio BS and using a diesel generator. Moreover, we have to be aware of the hazards of such advancement and take control before its too late. Therefore, we propose new logistic technology to save energy, avoid EMR eliminates carbon emission and power consumption. For improving the life quality in the smart city, tethered balloon technology can play a vital role in solving hazardous challenges of radio BS via mitigation biological, heating, and effects of ecological hazardous. Tethered balloon technology can play a vital solution to reduce health hazard and environmental hazards through energy-efficient radio equipment and solar system. It is the promise and emergent technology for green communication and could improve human life quality and make it more comfortable. Hence, it becomes possible solution to reduce the ill effects of cell tower radiation.

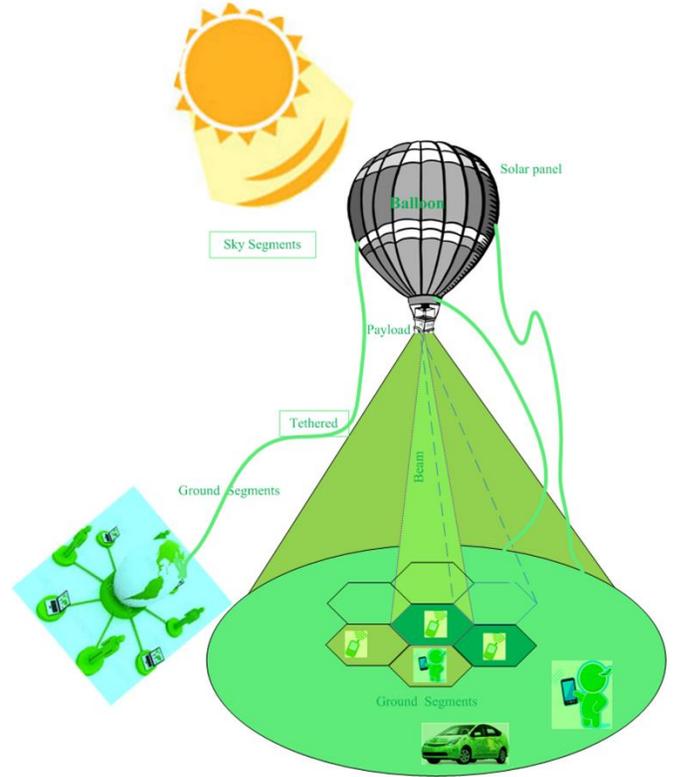

Fig. 3: Tethered balloon technology for green communication

The proposed network architecture is shown in Fig. 3. It consists of space and ground segments. Space segments include the payload, balloon, solar energy, and tether. The payload comprises remote radio unit (RRU) and antennas for mobile wireless communications. The proposed green broadband communication is anticipated to provide access capacity in the range of the underlying wireless access technology capacities

WiMAX. Also, a global positioning system (GPS) can be used for measuring the balloon height as well as it can be deployed to monitor the status of the floating platform Fig. 3. Tethered balloon technology plays a vital role in improving life quality due to improving green communication and reducing the impacts of traditional communication technologies.the most important of tethered balloon is the capability of delivering better broadband communication service, low cost, easy to deploy, could be lunched to different altitudes.Tether is the connecting link between the balloon and other networks on the ground. Fiber should be used inside the tether to connect the payload to the ground terminals for delivering broadband services in high capacity and very high speed. Furthermore, Balloon is restrained by the tether to maintain the balloon stable in position. On the other hand, the ground segment includes main unit (MU) of BS, sub-network, and mobile user.Briefly, tethered is used for several function i.e., delivering broadband services, restrict balloon to the ground and make it stable, and recharge the payload devices by energy in the case of solar panel is failed. MU is the system, which interfaces for 2 Mbps transmission links to the existing core network in public switched telephone networks (PSTN). The ground station acts as a hub for the network and provides service to customers within the cell radius. Furthermore, designing the space of balloon with solar panel will play vital role in reducing the energy used from ground to supply the payload.

## VI. RESULT AND DISCUSSIONS

The harmful of EMR to the human body is depending on the radiated power density and the distance from the transmitter. The first scenario, the distance of near field is considered 25m from the launch point of a tethered balloon. The balloon heights are 150m and 200m, as shown in Fig. 4 and Fig. 5. The power density from Balloon at these heights are shown in the maximum power density $1.98 \times 10^{-3} W/m^2$ when the height of the tethered balloon is 200m. Also, the maximum power density is $3.537 \times 10^{-3} W/m^2$ when the height of tethered balloon is 150m. Therefore, the power density decreased when the height increased, and vice versa.

In the second scenario, the distance from the tethered balloon to the user at the ground is considered with a different altitude from 200m to 400m. The results show that the distance from the balloon to the user increase by increasing the balloon height. Therefore, increasing the distance leads to decrease the power density gradually from maximum 2.5nW/m, when the user at the center coverage area of the tethered balloon as shown in Fig. 6. Furthermore, the radiation power also decreases with increases the distance from the balloon as shown in Fig. 7. The received power at the destination is shown also in Fig. 8.

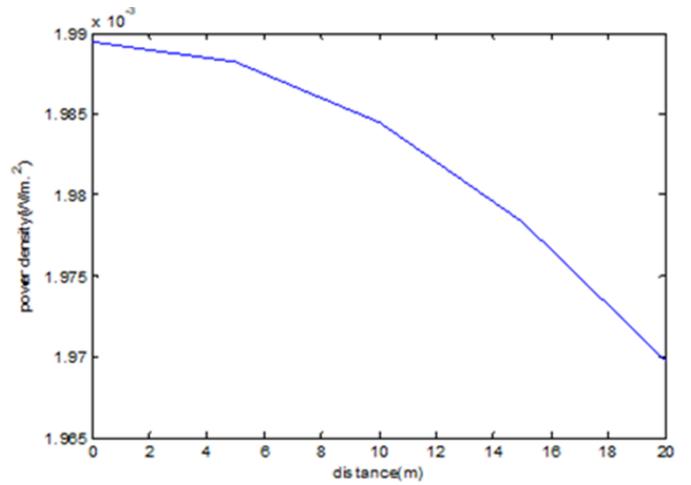

Fig. 5: Power density at the height of 200m

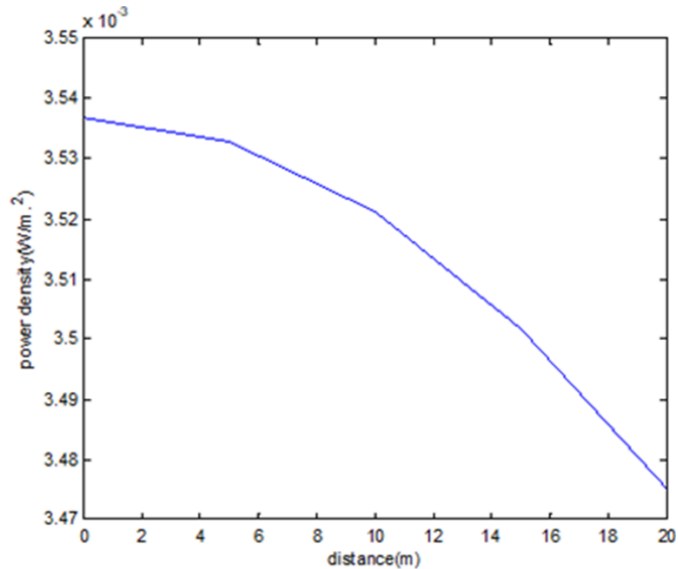

Fig. 4: Power density at the height of 150m

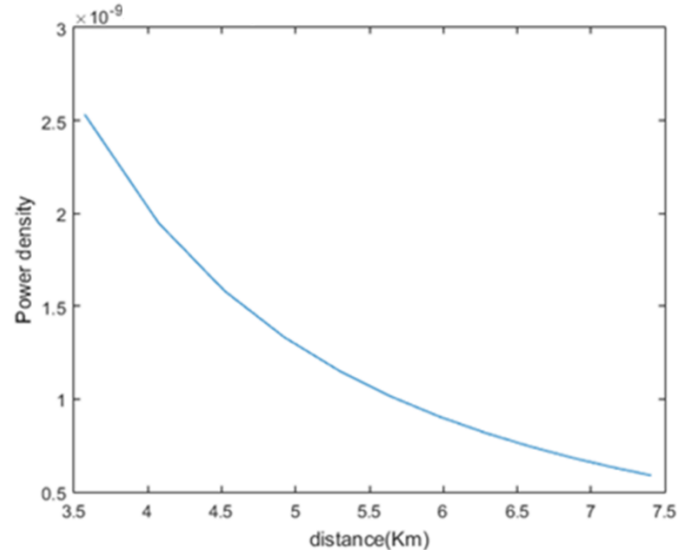

Fig. 6: Power density from ballon of different altitude

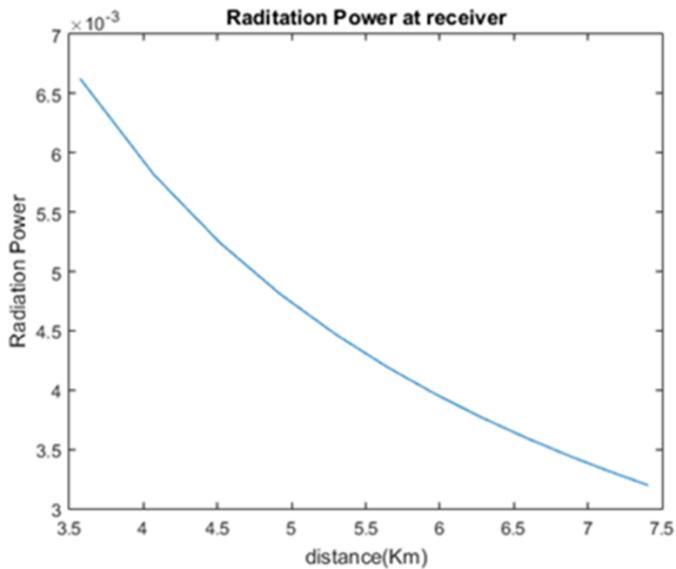

Fig. 7: Radiation Power from different distance

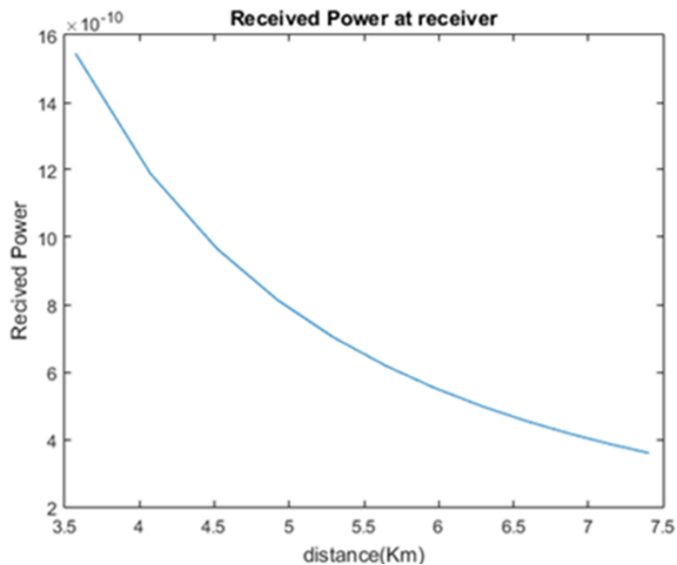

Fig. 8: Received power from Balloon in different heights

## VII. CONCLUSION

Tethered balloon is the newest and promising technology which can be used efficiently for green communication and healthy environments. Tethered balloon can carry the BS in its payload instead of increasing the height of BS over the house, which will affect the health of people who are living in the same house or nearest to BS. Power density values from a different height of tethered balloon are observed. Increasing the distance between radio BS and receiver leads to a decrease the values of power density at near field. The power density, radiation power, and received power are measured with different distances from Balloon at different altitudes. Using the proposed technique leads to an increase in the coverage area, avoid intense EMR even if there are many operators, reduce $CO_2$ emission as this system could be a solarpowered system. The results are shown that tethered balloon is efficient and optimal broadband communication without harming human life quality and society even if more than one operator shares the same tethered balloon. Thus, tethered balloon plays a vital solution to protect us from the harmful effects of the intense exposure to EMF, carbon emission, and environmental hazards. In the future, the tethered balloon may use for greening IoT for being considered as fog computing for improving energy efficiency and real-time service delivery.